\documentclass[conference]{IEEEtran}
\IEEEoverridecommandlockouts
\usepackage{amsmath,amssymb,amsfonts}
\usepackage{graphicx}
\usepackage{textcomp}
\usepackage{xcolor}

\usepackage{amsmath,amsthm,amssymb}
\usepackage{enumerate,tikz}

\newtheorem{theorem}{Theorem}

\newtheorem{lemma}[theorem]{Lemma}

\theoremstyle{definition}

\DeclareMathOperator*{\argmin}{arg\,min}
\renewcommand{\hat}{\widehat}
\renewcommand{\tilde}{\widetilde}
\renewcommand{\epsilon}{\varepsilon}

\newcommand{\commentout}[1]{}

\def\supp{\text{supp}}

\def\<{\langle}
\def\>{\rangle}
\def\({\Big(}
\def\){\Big)}
\def\calA{\mathcal{A}}
\def\C{\mathbb{C}}

\def\calF{\mathcal{F}}
\def\calZ{\mathcal{Z}}

\def\calIM{\mathcal{I}^M}
\def\calIm{\mathcal{I}^m}

\def\T{\mathbb{T}}

\def\Z{\mathbb{Z}}

\begin{document}

\title{High-performance quantization for spectral super-resolution}

\author{\IEEEauthorblockN{C.~Sinan G\"unt\"urk}
\IEEEauthorblockA{\textit{Courant Institute of Mathematical Sciences} \\
\textit{New York University}\\
New York, NY, USA \\
gunturk@cims.nyu.edu}
\and
\IEEEauthorblockN{Weilin Li}
\IEEEauthorblockA{\textit{Courant Institute of Mathematical Sciences} \\
\textit{New York University}\\
New York, NY, USA \\
weilinli@cims.nyu.edu}
}

\maketitle

\begin{abstract}
We show that the method of distributed noise-shaping beta-quantization offers superior performance for the problem of spectral super-resolution with quantization whenever there is redundancy in the number of measurements. More precisely, if the (integer) oversampling ratio $\lambda$ is such that $\lfloor M/\lambda\rfloor - 1\geq 4/\Delta$, where $M$ denotes the number of Fourier measurements and $\Delta$ is the minimum separation distance associated with the atomic measure to be resolved, then for any number $K\geq 2$ of quantization levels available for the real and imaginary parts of the measurements, our quantization method guarantees reconstruction accuracy of order $O(\lambda^{3/2} K^{- \lambda/2})$, up to constants which are independent of $K$ and $\lambda$. In contrast, memoryless scalar quantization offers a guarantee of order $O(K^{-1})$ only.
\end{abstract}


\section{Introduction}

Analog-to-digital conversion is inherently lossy. It typically consists of two stages, sampling and quantization, where the sampling stage produces a stream of scalar samples and the quantization stage replaces each sample with an element of a discrete set, called the quantization alphabet. Ideally the sampling stage is lossless (or the distortion is negligible) so that the distortion is only (or primarily) caused by quantization. 

The naive method of quantization is to round each scalar measurement to the nearest available level in the quantization alphabet; this is known as memoryless scalar quantization (MSQ). While MSQ has many advantages in hardware implementation (e.g. simplicity and robustness), its rate-distortion performance is highly suboptimal when the sampling map is redundant, meaning it collects more measurements than the minimal number needed for perfect reconstruction. This is due to the fact that a redundant sampling system increases the ambient dimension of the measurement vectors but not the intrinsic dimension of the manifold on which these vectors lie, and therefore, memoryless scalar quantized vectors, lying in a small neighborhood of this manifold, occupy only an asymptotically vanishing portion of the rectangular lattice of available quantization vectors. More efficient quantization methods achieve improved rate-distortion performance by utilizing some (or all) of the remaining quantization vectors as well. Noise-shaping quantizers (such as $\Sigma\Delta$ modulators) fall into this category.

This paper provides a new approach of quantizing non-harmonic Fourier measurements for the spectral super-resolution problem. Super-resolution has received considerable attention in the past several years (e.g. \cite{candes2013super,candes2014towards}). 
The goal is to accurately estimate an unknown discrete measure 
\begin{equation}
 \label{eq:mu}
\mu=\sum_{j=1}^S a_j\delta_{t_j} 
\end{equation}
defined on $\T:=[0,1)$, from its noisy samples 
$$\tilde y_k = y_k + z_k, \qquad k=0,\dots,M-1,$$ 
where 
\begin{equation}
\label{eq:hatmu}
y_k := \hat\mu(k)
:=\int_0^1 e^{-2\pi ik t}\ d\mu(t)
=\sum_{j=1}^S a_j e^{2\pi i k t_j}
\end{equation}
is the $k$-th Fourier coefficient of $\mu$ and the unknown noise vector $z:=(z_k)_0^{M-1}$ satisfies $\|z\|_2\leq \epsilon$ for some known $\epsilon > 0$.
We emphasize that the total number of spikes $S$, the amplitudes $a\in\C^S$, and the support set $T=\{t_j\}_{j=1}^S$ are unknown. 


This problem is ill-conditioned if there are points in $T$ that are too close to one another (e.g. \cite{donoho1992superresolution,demanet2015recoverability,li2017stable}). However, assuming a lower bound to their minimum separation, it has been shown that $\mu$ can be recovered from its measurements in a robust way when $M$ is sufficiently large, meaning that the reconstruction error, when measured in a suitable metric, is controlled by the noise energy in a graceful manner, and typically linearly (e.g. \cite{fernandez2013support}). 

Since no structure is assumed on the noise, robustness also becomes the key property that allows for quantization. In particular, it guarantees that sufficiently high-resolution quantizers will always produce sufficiently high quality approximations. However, the question of achievable limits of quantization accuracy is open. The answer depends on the interplay between the given (fixed) parameters, such as the number of measurements, the number of quantization levels per measurement, and the minimum separation distance of the measures of interest, as well as the quantization method which itself is a design parameter. 

For simplicity, we will assume in this paper that quantization is the only source of perturbation. Other (generally uncontrolled) sources of perturbations can be incorporated into quantization as well; any such perturbations typically provide a noise floor. We assume that the real and imaginary parts of each Fourier measurement is replaced by an element of a (signal-independent) quantization alphabet $\calA$ with $K$ levels. 

The linear dependence of the reconstruction error on the noise energy implies that with simple rounding, i.e. MSQ, it is straightforward to achieve reconstruction accuracy of order $O(K^{-1})$. However, since the noise energy is measured in $\ell_2$, its bound $\epsilon$ grows with the number of measurements, and therefore it is not even clear if there is any advantage of using any additional measurements. For example, the popular reconstruction method total-variation minimization (TV-min, see Section \ref{sec:quan}) appears to be indifferent to oversampling in practice. 

Our main result in this paper is that there is an alternative quantization method, called the distributed noise-shaping $\beta$-encoder (or $\beta$-quantization in short) which, together with an accompanying alternative recovery method derived from TV-min, is able to exploit any available redundancy. More precisely, 
if the (integer) oversampling ratio $\lambda$ is such that $\lfloor M/\lambda\rfloor - 1\geq 4/\Delta$, where $M$ denotes the number of Fourier measurements and $\Delta$ is the minimum separation distance associated with $\mu$, then for any number $K$ of quantization levels, our quantization method guarantees reconstruction accuracy of order $O(\lambda^{3/2} K^{-\lambda/2})$, up to constants which are independent of $K$ and $\lambda$. In principle our method can work with other robust recovery methods, too. 

The paper is organized as follows. Section \ref{sec:TV} reviews the TV-min method and discusses MSQ for spectral super-resolution. In Section \ref{sec:TV}, we introduce the proposed quantization method and the main ingredients needed for its error performance analysis for spectral super-resolution, which is done in Section \ref{sec:SR}. Finally, we provide a sample numerical experiment to demonstrate the practical performance of our proposed quantization method in Section \ref{sec:num}.

\section{A review of TV-min for super-resolution and MSQ}
\label{sec:TV}

There are a number of robust recovery algorithms for super-resolution. For convenience and concreteness, we will focus on TV-min, also known under the name BLASSO (Beurling Lasso). 

Let us denote by $\calF_M$ the operator which maps the measure $\mu$ to $(\hat \mu(k))_0^{M-1}$, i.e. with our notation of \eqref{eq:hatmu}, we have $y=\calF_M\mu$, where $y:=(y_k)_0^{M-1}$. Given noisy data $\tilde y\in\C^{M}$ and a bound $\epsilon > 0$ on the noise, the TV-min algorithm outputs an estimate $\tilde \mu$ of $\mu$ given by 
\begin{equation} \label{eq:TV-min}
\tilde\mu 
:=\argmin\{\|\nu \|_{TV}\colon 
\|\calF_M \nu-\tilde y\|_2 \leq \epsilon\}.
\end{equation}
This is a convex program whose feasibility is guaranteed by the assumption $\|y - \tilde y\|_2 \leq \epsilon$. The solution may not be unique, but it is known that there is at least one minimizer which is a discrete measure which we will identify with $\tilde \mu$. 
(See \cite{candes2014towards,candes2013super,duval2015exact,benedetto2018super} for this and other results.) 

The performance of TV-min depends on the minimum separation of the measure, defined as
\begin{equation}
\label{eq:minsep}
\Delta(\mu)
:=\min_{s\not=t, s,t\in\supp(\mu)} |s-t|_\T. 
\end{equation}
Here, $|s-t|_\T:=\min_{n \in \Z}~|s-t-n|$ is the ``wrap-around'' metric on $\T$. If the number of samples $M$ is sufficiently large so that 
\begin{equation}
\Delta(\mu)\geq \frac{4}{M-1},
\end{equation}
then $\tilde \mu$
provides an accurate estimate of $\mu$ in the following sense: any spurious spikes in $\tilde \mu$ are smaller than $O(\epsilon)$, the remaining spikes in $\tilde\mu$ are within $O(1/M)$ of the true spikes, and the recovered amplitudes are within $O(\epsilon)$ of the true ones. More precisely, given the representation
\begin{equation}\label{eq:tildemu}
\tilde \mu=\sum_{k=1}^{\tilde S} \tilde a_k \delta_{\tilde t_k},
\end{equation}
along with the ``neighborhood'' index sets
\begin{equation}\label{eq:Ij}
\calIM_j:=\big \{k: |\tilde t_k-t_j|_\T \leq 2\cdot 0.1649\,(M-1)^{-1} \big \},~~ j=1,\dots,S,
\end{equation}
and the residual index set 
\begin{equation}\label{eq:I0}
\calIM_0 := \{1,\dots,\tilde S\}\setminus \bigcup_{j=1}^S \calIM_j,
\end{equation}
the following bounds are guaranteed:
\begin{eqnarray}
 \big| a_j -\sum_{k \in \calIM_j} \tilde a_k \big| & \leq & C_1 \epsilon,~~~j=1,\dots,S, \label{eq:rec1} \\
 \sum_{k \in \calIM_j} |\tilde a_k|\,|t_j - \tilde t_k|_\T^2 & \leq & C_2 M^{-2}\epsilon, ~~~j=1,\dots,S,  \label{eq:rec2} \\
 \sum_{k \in \calIM_0} |\tilde a_k | & \leq & C_3 \epsilon \label{eq:rec3}.
\end{eqnarray}
Additional details can be found in \cite[Theorem 1.2]{fernandez2013support}; see also \cite[Theorems 2.1 and 2.2]{azais2015spike} for related results. 

This result provides an immediate error bound for MSQ. For each integer $K \geq 2$, let $\calZ_K$ denote the $K$-term origin-symmetric arithmetic progression of integers with spacing $2$ and $\calA_K:= K^{-1}\calZ_K \subset (-1,1)$. It is then clear that for all $u \in [-1,1]$ there exists $q \in \calA_K$ such that $|u - q| \leq 1/K$. Assuming that $\|\mu\|_{TV}\leq 1$ so that $\|y\|_\infty \leq 1$, it follows that for each complex measurement $y_k$, there is an element $q_k \in \calA_K + i \calA_K$ (found by separately rounding the real and the imaginary parts of $y_k$ to elements of $\calA_K$) such that $|y_k - q_k| \leq \sqrt{2}/K$.
Consequently, we have $\|y - q\|_2 \leq \sqrt{2M}/K$. Setting $\tilde y = q$ and $\epsilon=\sqrt{2M}/K$ in \eqref{eq:TV-min} guarantees, in view of \eqref{eq:rec1}-\eqref{eq:rec3}, an overall reconstruction accuracy of $O(\sqrt{M}/K)$. 

We can produce a lower bound on the worst-signal reconstruction error of MSQ as follows: Even if we knew the support $T$ of $\mu$, where $|T| = S$, memoryless scalar quantization of the $M$ linear measurements of all possible coefficient vectors $a \in \C^S$ chosen from any fixed ball results in a partition of this ball using at most $O(MK)$ hyperplanes, and therefore into at most $(cMK/S)^{2S}$ cells. (Here $c$ is an absolute constant.) Consequently, there will always be a cell of diameter at least $O(S/MK)$ whose elements are all mapped to the same quantized vector. 

Therefore, suppressing the dependence on $M$, it follows that MSQ cannot offer error performance better than $O(K^{-1})$.

\section{Proposed quantization method}
\label{sec:quan}

Our proposed quantization approach in this paper is based on the general framework of distributed noise-shaping $\beta$-encoding developed in \cite{chou2016distributed}, \cite{chou2017distributed} (see also \cite{chou2013beta} and \cite{chou2015noise} for prior versions). However, the specialization for the spectral super-resolution problem requires some new choices and adaptations.

Let $\lambda \geq 1$ be an integer which should be thought of as a lower bound on the oversampling ratio. For the simplicity of discussion, we assume $M$ is divisible by $\lambda$ and set $m := M/\lambda$. Let $\beta>1$ be a parameter which shall be chosen later, and consider the $m\times M$ matrix $V:=V_{\beta}$
\begin{equation}
\label{eq:V}
V:=
\begin{bmatrix}
I_m &\beta^{-1}I_m &\cdots &\beta^{-\lambda+1} I_m
\end{bmatrix}, 
\end{equation}
where $I_m$ denotes the $m\times m$ identity matrix. Observe that 
\[
(Vy)_\ell
=\sum_{k=0}^{\lambda-1} \beta^{-k} y_{mk+\ell}
=\sum_{j=1}^S a_j w_j e^{-2\pi i \ell t_j}
\]
for $\ell=1,\dots,m$, where 
\begin{equation}\label{eq:w}
w_j:=
\frac{1-\beta^{-\lambda}e^{-2\pi i m\lambda t_j}}{1-\beta^{-1}e^{-2\pi i mt_j}}, ~~j=1,\dots,S. 
\end{equation}
In other words, we have the relation $Vy = \calF_m \mu_V$ where
\begin{equation}
\label{eq:muV}
\mu_V:=\sum_{j=1}^S b_j \delta_{t_j}
\quad\text{and}\quad b_j := a_j w_j. 
\end{equation}
Observe that $\mu$ and $\mu_V$ have identical supports, but different amplitudes. However, the weights $w_j$ satisfy 
\begin{equation}
\label{eq:weights}
\frac{1}{c_\beta}
\leq |w_j|
\leq c_\beta 
\quad\text{where}\quad
c_\beta := \frac{1+\beta^{-1}}{1-\beta^{-1}}.
\end{equation}

Let us define $H:=H_\beta$ to be the $M\times M$ matrix where 
\begin{equation}
\label{eq:H}
H_{j,k}:=
\begin{cases} 
1, & \mbox{if }j=k, \\
-\beta, & \mbox{if }j = k{+}m \mbox{ and } 1 \leq k \leq M{-}m. 
\end{cases}
\end{equation}

The following is a special case of \cite[Lemma 2]{chou2017distributed}:

\begin{lemma}
	Let $K\geq 2$ be an integer, and suppose the parameters $\alpha,\beta,\delta>0$ satisfy the inequality
	\begin{equation}
	\label{eq:parameter}
	\beta + \alpha\delta^{-1} \leq K, 
	\end{equation}
	and consider the quantization alphabet $\calA := \delta (\calZ_K+i\calZ_K)$. Then,
	for any $y\in\C^M$ with $\|y\|_\infty\leq\alpha$, there exists $q\in\calA^M$ and $u\in\C^M$ with $\|u\|_\infty\leq \sqrt{2} \delta$ satisfying the relationship
	\begin{equation}\label{eq:yqHu}
	y-q=H u.
	\end{equation}
\end{lemma}

The mapping $y \mapsto q$ implied by the above lemma can be implemented by means of a simple recursive algorithm. We omit the details and refer to \cite{chou2016distributed, chou2017distributed}.

The significance of $V$ and $H$ is that $VH$ is very small when $\beta$ or $\lambda$ is large. Indeed, as shown in \cite{chou2016distributed}, we have
$\|VH\|_{\infty\to 2} = \sqrt{m} \beta^{-\lambda+1}$. The immediate consequence is that 
\begin{equation}
\label{eq:Verror}
\|Vy-Vq\|_2
\leq \|VH\|_{\infty\to 2}\|u\|_\infty
\leq \sqrt{2m}\ \beta^{-\lambda+1} \delta.
\end{equation}

The above findings provide the core strategy of our proposed quantization and recovery method. First, we
note that for any $1< \beta < K$ and any $0 \leq \alpha < \infty$, there exists $\delta > 0$ such that the condition \eqref{eq:parameter} is satisfied. Hence the existence of the mapping $y \mapsto q$ (with a fixed quantization alphabet $\calA$) is guaranteed over any bounded set of inputs $y$. Next, 
recall that with $y = \calF_M \mu$, we have $Vy = \calF_m \mu_V$. Since $Vq$ is now a small perturbation of $Vy$, we can obtain a close approximation $\tilde \mu_V$ of $\mu_V$ by means of any robust super-resolution recovery method, such as the TV-min algorithm. Then, since $\mu$ and $\mu_V$ have identical supports, we can define an approximate recovery $\tilde \mu$ by means of approximate weights $\tilde w_j$ derived from the approximate support $\tilde T$ of $\tilde \mu_V$. 

Let us summarize the proposed quantization method. 

\par \noindent {\bf System parameters and assumptions:}
\begin{itemize}
 \item $M$ (number of Fourier measurements),
 \item $\alpha$ (upper bound on TV-norm of the input measures),
 \item $\Delta$ (lower bound on minimum separation distance),
 \item $M=\lambda m$, $\lambda \geq 1$, $m-1\geq 4/\Delta$, 
 \item $K$ (number of quantization levels),
 \item $\beta$ and $\delta$ such that \eqref{eq:parameter} holds.
\end{itemize}

\par \noindent {\bf Encoding (quantization) stage:}
\begin{itemize}
 \item Input to quantizer: $y$ such that $\|y\|_\infty \leq \alpha$,
 \item $V$ and $H$ defined via \eqref{eq:V} and \eqref{eq:H},
 \item Quantization alphabet: $\calA:= \delta(\calZ_K + i \calZ_K)$,
 \item Output of quantizer: $q\in \calA^M$ such that $\|Vy - Vq\|_2 \leq  \sqrt{2m}\ \beta^{-\lambda+1} \delta$.
\end{itemize}

\par \noindent {\bf Decoding (recovery) stage:}
\begin{itemize}
 \item Input to decoder: $q$,
 \item Compute a minimum TV-norm measure $\tilde\mu_V$ of the form $\sum_{k=1}^{\tilde S} \tilde b_k \delta_{\tilde t_k}$ satisying $\|\calF_m \tilde\mu_V-Vq\|_2 \leq \epsilon_V$ where 
 $$\epsilon_V:=\sqrt{2m}\ \beta^{-\lambda+1} \delta;$$ 
 abort if it cannot be found (e.g. invalid measurements),
 \item Set $$\tilde w_k = \frac{1-\beta^{-\lambda}e^{-2\pi i m\lambda \tilde t_k}}{1-\beta^{-1}e^{-2\pi i m \tilde t_k}},~~k=1,\cdots,\tilde S,$$
 \item Output of decoder: $\tilde \mu = \sum_{k=1}^{\tilde S} \tilde a_k \delta_{\tilde t_k}$ with $\tilde a_k:= \tilde b_k/\tilde w_k$.
\end{itemize}

\section{Error analysis}
\label{sec:SR}

Let us start by noting that when the input to the quantizer $y$ equals
$\calF_M\mu$ for some $\mu$ of the form \eqref{eq:mu} with $\Delta(\mu) \geq \Delta$ and $\|\mu\|_{TV} \leq \alpha$, then the decoder will always output a measure $\tilde \mu$, thanks to the fact that $Vy = \calF_m \mu_V$ where $\mu_V$ is defined by \eqref{eq:muV} which guarantees that $\mu_V$ is a feasible measure for the TV-min program.

Let us now proceed to find an error bound for $\tilde \mu$. We start by comparing $\tilde \mu_V$ to $\mu_V$.
With the error bounds of the general TV-min method reviewed in Section \ref{sec:TV}, we have
\begin{eqnarray}
 \big| b_j -\sum_{k \in \calIm_j} \tilde b_k \big| & \leq & C_1 \epsilon_V,~~~j=1,\dots,S, \label{eq:brec1} \\
 \sum_{k \in \calIm_j} |\tilde b_k|\,|t_j - \tilde t_k|_\T^2 & \leq & C_2 m^{-2}\epsilon_V, ~~~j=1,\dots,S,  \label{eq:brec2} \\
 \sum_{k \in \calIm_0} |\tilde b_k | & \leq & C_3 \epsilon_V \label{eq:brec3}.
\end{eqnarray}
where the index sets $\calIm_j$, $j=0,\dots,S$ are as in \eqref{eq:Ij} and \eqref{eq:I0}, only for $m$ measurements. Note that for all $j \in \{1,\dots,S\}$,
\begin{equation}\label{eq:step1}
\Big| a_j{-}\sum_{k \in \calIm_j} \tilde a_k \Big| 
\leq \frac{1}{|w_j|}\Big| b_j{-}\sum_{k \in \calIm_j} \tilde b_k \Big| {+}  \sum_{k \in \calIm_j} |\tilde b_k|\Big| \frac{1}{w_j} -\frac{1}{\tilde w_k}  \Big|. 
\end{equation}
With \eqref{eq:weights} and \eqref{eq:brec1}, the first term is bounded by $c_\beta C_1 \epsilon_V$. For the second term, we note that 
\begin{equation}\label{eq:boundinvw}
 	\Big| w_j^{-1} -\tilde w_k^{-1}  \Big| \leq m\, C_{\beta, \lambda}\, |t_j-\tilde t_k|_\T
\end{equation}
where $C_{\beta,\lambda}$ stands for the Lipschitz constant of the map 
$$t \mapsto \frac{1-\beta^{-1}e^{-2\pi i t}}{1-\beta^{-\lambda}e^{-2\pi i \lambda t}}, ~~~t \in \T.$$ It can be shown $C_{\beta,\lambda} \leq 4\pi \lambda \beta (\beta-1)^{-2}$.

Using \eqref{eq:boundinvw} and Cauchy-Schwarz, we see that the second term in \eqref{eq:step1} is bounded by
$$
m\,C_{\beta,\lambda} \Big (\sum_{k \in \calIm_j}|\tilde b_k|\ |t_j-\tilde t_k|^2_\T \Big)^{1/2}
\Big(\sum_{k \in \calIm_j} |\tilde b_k|\Big)^{1/2}
$$
Note that $\|\tilde b\|_1\leq c_\beta \alpha$ since $\|\tilde b\|_1=\|\tilde \mu_V\|_{TV}$ and
\[
\|\tilde \mu_V\|_{TV} \leq
\|\mu_V\|_{TV} 
	\leq \|w\|_\infty \|a\|_1
	\leq c_\beta \|\mu\|_{TV}
	\leq c_\beta \alpha. 
\]
Hence, with \eqref{eq:brec2} we deduce
\begin{equation}
  \label{eq:term2}
  \sum_{k \in \calIm_j} |\tilde b_k|\ \Big| \frac{1}{w_j} -\frac{1}{\tilde w_k}  \Big|
  \leq C_{\beta,\lambda} \sqrt{c_\beta \alpha} \sqrt{C_2 \epsilon_V}.
\end{equation}
Injecting \eqref{eq:term2} into \eqref{eq:step1} we have
\begin{equation}
\label{eq:amplitudes}
\Big| a_j-\sum_{k \in \calIm_j} \tilde a_k \Big| 
\leq c_\beta C_1 \epsilon_V + C_{\beta,\lambda} \sqrt{c_\beta \alpha} \sqrt{C_2 \epsilon_V}.
\end{equation}
Finally, we also have
\begin{eqnarray}
 \sum_{k \in \calIm_j} |\tilde a_k|\,|t_j - \tilde t_k|_\T^2 & \leq & c_\beta C_2 m^{-2}\epsilon_V,  \label{eq:arec2} \\
 \sum_{k \in \calIm_0} |\tilde a_k | & \leq & c_\beta C_3 \epsilon_V \label{eq:arec3}.
\end{eqnarray}
which follow readily from \eqref{eq:weights}, \eqref{eq:brec2} and  \eqref{eq:brec3}.

At this point, we note the following elementary fact: For any $K\geq 2$, setting $\beta:= K(\lambda+1)/(\lambda+2)$ and $\delta:=(\lambda+2)\alpha/K$ results in $\beta + \alpha\delta^{-1} = K$ and $\delta \beta^{-\lambda+1} < \mathrm{e} \alpha (\lambda+1) K^{-\lambda}$.
(See, e.g. \cite[Lemma 3.2]{chou2016distributed} and \cite[Lemma 1]{chou2017distributed}.)
This choice of parameters results in 
$\epsilon_V \leq \mathrm{e}\alpha\sqrt{2m}(\lambda+1) K^{-\lambda}$.
Furthermore it is readily seen that $\beta \geq 4/3$, $c_\beta \leq 7$ and $C_{\beta,\lambda} \leq 12\pi \lambda$.
Hence it follows from \eqref{eq:amplitudes}, \eqref{eq:arec2} and \eqref{eq:arec3} that $\mu$ is approximated by $\tilde \mu$ up to resolution $O(\sqrt{M} \lambda^{3/2} K^{-\lambda/2})$.
\section{Numerical results}
\label{sec:num}

\begin{figure}
	\caption{Average reconstruction error of MSQ and $\beta$-quantization. \label{fig:experiment}}
	\centering
	\includegraphics[width=0.48\textwidth]{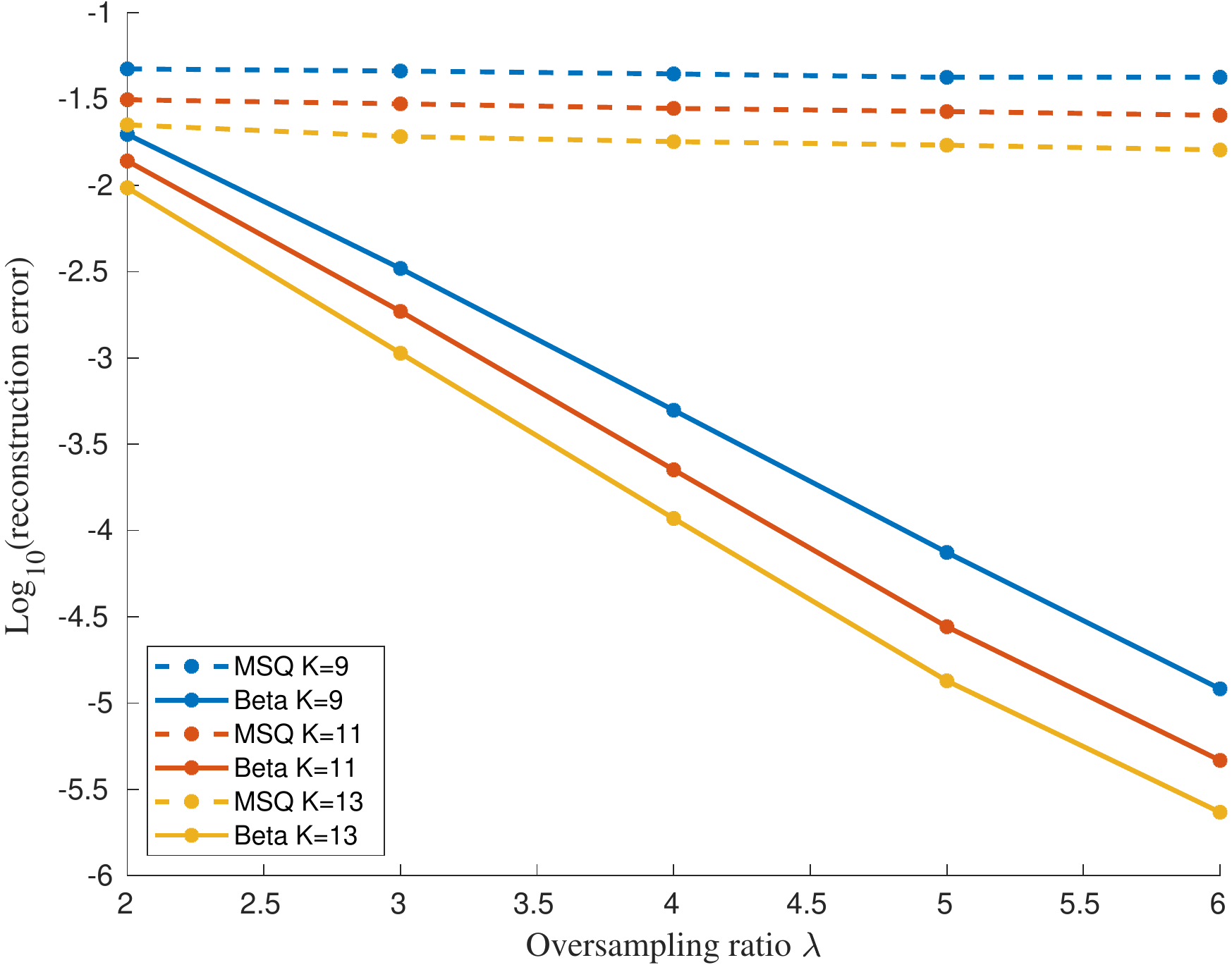}
\end{figure}

We compare the reconstruction error, quantified by the term on the left hand side of \eqref{eq:amplitudes}, when the Fourier samples are quantized using our proposed beta-quantization versus MSQ. More specifically, we set $\Delta=1/10$ and we randomly select a measure $\mu$ such that $\Delta(\mu)\geq \Delta$; the amplitudes are chosen uniformly at random and normalized to have unit $\ell^1$ norm. For various choices of $\lambda$ and $K$, we quantize the Fourier measurements using both MSQ and $\beta$-quantization. Figure \ref{fig:experiment} displays the reconstruction error as a function of $\lambda$, averaged over 110 trials. The experiment validates our theoretical results and also shows that performance of MSQ is suboptimal in the over-sampling regime.

\bibliography{QuanSRbib}

\begin{thebibliography}{10}

\bibitem{azais2015spike}
Jean-Marc Aza{\"i}s, Yohann De~Castro, and Fabrice Gamboa.
\newblock Spike detection from inaccurate samplings.
\newblock {\em Applied and Computational Harmonic Analysis}, 38(2):177--195,
  2015.

\bibitem{benedetto2018super}
John~J. Benedetto and Weilin Li.
\newblock Super-resolution by means of {B}eurling minimal extrapolation.
\newblock {\em Applied and Computational Harmonic Analysis}, 2018.

\bibitem{candes2013super}
Emmanuel~J. Cand{\`e}s and Carlos Fernandez-Granda.
\newblock Super-resolution from noisy data.
\newblock {\em Journal of Fourier Analysis and Applications}, 19(6):1229--1254,
  2013.

\bibitem{candes2014towards}
Emmanuel~J. Cand{\`e}s and Carlos Fernandez-Granda.
\newblock Towards a mathematical theory of super-resolution.
\newblock {\em Communications on Pure and Applied Mathematics}, 67(6):906--956,
  2014.

\bibitem{chou2013beta}
Evan Chou.
\newblock {\em Beta-duals of frames and applications to problems in
  quantization}.
\newblock PhD thesis, New York University, 2013.

\bibitem{chou2016distributed}
Evan Chou and C.~Sinan G{\"u}nt{\"u}rk.
\newblock Distributed noise-shaping quantization: {I}. {B}eta duals of finite
  frames and near-optimal quantization of random measurements.
\newblock {\em Constructive Approximation}, 44(1):1--22, 2016.

\bibitem{chou2017distributed}
Evan Chou and C.~Sinan G{\"u}nt{\"u}rk.
\newblock Distributed noise-shaping quantization: {II}. {C}lassical frames.
\newblock In {\em Excursions in Harmonic Analysis, Volume 5}, pages 179--198.
  Springer, 2017.

\bibitem{chou2015noise}
Evan Chou, C.~Sinan G{\"u}nt{\"u}rk, Felix Krahmer, Rayan Saab, and
  {\"O}zg{\"u}r Y{\i}lmaz.
\newblock Noise-shaping quantization methods for frame-based and compressive
  sampling systems.
\newblock In {\em Sampling theory, a renaissance}, pages 157--184. Springer,
  2015.

\bibitem{demanet2015recoverability}
Laurent Demanet and Nam Nguyen.
\newblock The recoverability limit for superresolution via sparsity.
\newblock {\em arXiv preprint arXiv:1502.01385}, 2015.

\bibitem{donoho1992superresolution}
David~L. Donoho.
\newblock Superresolution via sparsity constraints.
\newblock {\em SIAM Journal on Mathematical Analysis}, 23(5):1309--1331, 1992.

\bibitem{duval2015exact}
Vincent Duval and Gabriel Peyr{\'e}.
\newblock Exact support recovery for sparse spikes deconvolution.
\newblock {\em Foundations of Computational Mathematics}, 15(5):1315--1355,
  2015.

\bibitem{fernandez2013support}
Carlos Fernandez-Granda.
\newblock Support detection in super-resolution.
\newblock In {\em Proceedings of the 10th International Conference on Sampling
  Theory and Applications}, pages 145--148, 2013.

\bibitem{li2017stable}
Weilin Li and Wenjing Liao.
\newblock Stable super-resolution limit and smallest singular value of
  restricted fourier matrices.
\newblock {\em arXiv preprint arXiv:1709.03146}, 2017.

\end{thebibliography}
\bibliographystyle{plain}

\end{document}